\documentclass[aps,floatfix]{revtex4}
\usepackage{graphicx}
\usepackage{epsfig}
\usepackage{amsmath}
\usepackage{graphics}
\usepackage{latexsym}
\begin{document}

\newcommand{\nc}{\newcommand}
 
\newcommand{\be}{\begin{equation}}
\newcommand{\ee}{\end{equation}}
\newcommand{\bea}{\begin{eqnarray}}
\newcommand{\eea}{\end{eqnarray}}
\newcommand{\R}{\vec{R}}
\newcommand{\bc}{\begin{center}}
\newcommand{\ec}{\end{center}}

\title{Star Polymers Confined in a Nanoslit:\\ A Simulation Test of Scaling and
Self-Consistent Field Theories}
\author{J. Paturej$^{1,2}$, Andrey Milchev$^{3,4}$, S. A. Egorov$^{5}$, and
Kurt Binder$^3$}
\affiliation{$^1$ Department of Chemistry, University of North
Carolina, Chapel Hill, North Carolina 27599-3290, USA \\ $^2$ Institute of
Physics, University of Szczecin, Wiekopolska 15, 10451 Szczecin, Poland\\
$^3$Institute of Physics, Johannes Gutenberg-Universit\"at Mainz, Staudinger Weg
7, 55099 Mainz, Germany \\ $^4$ Institute of Physical Chemistry, Bulgarian
Academy of Sciences, Sofia 1113, Bulgaria \\ $^5$ Department of Chemistry,
University of Virginia, Charlottesville, Virginia 22901, USA}

\begin{abstract}
The free energy cost of confining a star polymer where $f$ flexible polymer
chains containing $N$ monomeric units are tethered to a central unit in a slit
with two parallel repulsive walls a distance $D$ apart is considered, for good
solvent conditions. Also the parallel and perpendicular components of the
gyration radius of the star polymer, and the monomer density profile across the
slit are obtained. Theoretical descriptions via Flory theory and scaling
treatments are outlined, and compared to numerical self-consistent field
calculations (applying the Scheutjens-Fleer lattice theory) and to Molecular
Dynamics results for a bead-spring model. It is shown that Flory theory and
self-consistent field (SCF) theory  yield the correct scaling of the parallel
linear dimension of the star with $N$, $f$ and $D$, but cannot be used for
estimating the free energy cost reliably. We demonstrate that the same problem
occurs already for the confinement of chains in cylindrical tubes. We also
briefly discuss the problem of a free or grafted star polymer interacting with a
single wall, and show that the dependence of confining force on the
functionality of the star is different for a star confined in a
nanoslit and a star interacting with a single wall, which is due to
the absence of a symmetry plane in the latter case. 
\end{abstract}

\maketitle

\section{Introduction}

The physical properties and geometric conformations of macromolecules are
strongly affected when these polymers are confined in nanoscopically thin slits
or tubes, with linear dimensions in between the size of a monomeric unit and the
size of a free macromolecule in solution
\cite{1,MD_PJdG,3,4,5,6,7,8,9,10,11,12,13,14,15,16,17}. For linear
macromolecules, this problem has been considered extensively in the literature,
but much less attention has been devoted to the related problem of confining
polymers with a more complex chemical architecture, such as star polymers with
$f$ arms \cite{halperin87,19,19'}, dendrimers \cite{19}, randomly branched
polymers, etc. Such questions are important in various contexts, such as
chromatographic separations, colloidal stabilization (recall that a nanocolloid
coated with a grafted polymeric brush layer in many respects has properties
closely related to a star polymer \cite{20}), preparation of stimuli-responsive
nanomaterials, biomolecules constrained by cell membranes, etc.  Understanding
such problems first for isolated macromolecules and confining purely repulsive
walls is a prerequisite before one can address many related relevant problems
such as adsorption of such macromolecules at the confining surfaces \cite{21,CNL,CNL1},
interactions between confined polymers under various conditions
\cite{19',22,23,Egorov}, etc.

In the present work, we examine the generic problem of star polymers under
nanoconfinement, focusing on the simple limit where the number of arms $f$ is
not excessively large \cite{halperin87,24,25,25'}, and hence the physical
effects of crowding too many arms near the star center need not be considered.
As a first step, we formulate in Sec.~\ref{sec_theory} a crude version of the
Flory theory (where Gaussian entropic elasticity is combined with a mean-field
treatment of monomer-monomer-repulsions \cite{26}). This treatment predicts
correctly the dependence of linear dimensions on the number of arms $f$, their
chain length $N$, and the width of the slit $D$ (or pore diameter,
respectively). However, the corresponding free energies disagree with the
pertinent predictions of the scaling theory \cite{24,25}, which is based on the
simple blob concepts pioneered by de~Gennes and Daoud \cite{1,MD_PJdG}, or Daoud
and Cotton \cite{24}. As is well known, the success of the Flory theory for
chain linear dimensions of a single chain is due to the cancellation
of errors, although the precise reason for this cancellation is yet unknown~\cite{1}.

The free energy of a single chain under good solvent conditions according to
Flory theory is predicted to vary like $N^{1/5}$ with chain length $N$ while
scaling arguments imply that it is independent of $N$ (of the order of thermal
energy $k_BT$ \cite{1}). Thus this partial failure of Flory theory for stars is
not unexpected. Furthermore, the Flory treatment cannot describe the smooth
crossover to the case of unconfined star when $D$ gets comparable to the
diameter of a free star. For the description of this crossover, the lattice
formulation \cite{fleer93} of the self-consistent field theory
(Sec.~\ref{sec_SCFT}) clearly is a more powerful tool, although it shares
difficulties with the Flory theory with respect to the scaling properties of the
free energy. These problems of self-consistent field theory are also elucidated
by presenting results for the simpler problem of confining linear chains in
cylindrical tubes, \cite{Hsu2,Klushin} Section \ref{subsec_SCF_MC}. 
Molecular Dynamics simulations
\cite{28,29,30,31} (Sec.~\ref{sec_results}), on the other hand, can reproduce
the scaling predicted by the blob picture. With respect to local effects near
the confining walls, we discuss the use of the extrapolation length concept, and
find it similarly useful as for confined linear polymers
\cite{11,17}.  Finally, we also
consider the problem of a star polymer interacting with a
single wall, and show that the dependence of confining force on the
functionality of the star polymer is different for a star confined in a
nanoslit and a star interacting with a single wall, which is due to
the absence of a symmetry plane in the latter case. 
Section \ref{sec_concl} then summarizes our conclusions.

\section{FLORY THEORY OF FREE AND CONFINED STAR POLYMERS}
\label{sec_theory}
 
\subsection{Unconfined (Free) Stars}

The statistical mechanics of star polymers in good solvent has been treated by
many elaborate theories, from self-consistent minimization of intramolecular
interactions \cite{Allegra} to renormalization group methods
\cite{Miyake,Douglas,Vlahos,Duplantier1,Duplantier2,Ohno}. However, to provide
a first overview it is nevertheless useful to discuss the simple Flory theory. 

According to Flory theory, the free energy of a polymer in the presence of
excluded volume interactions is written as a sum of two terms, the first one
$({F}_{e \ell})$ corresponds to entropic elasticity of the chain, the
second to the (pairwise) interactions between the effective monomers
$(F_{\rm int})$, see Grosberg and Khokhlov \cite{GK},
\begin{equation}
 F = F_{e\ell} + F_{\rm int}
\end{equation}

For a single chain, $F_{e \ell}$ (in units of temperature) is $R^2 / a^2
N$, with $a$ the monomer size, $N$ the number of monomers. If $F_{\rm
int}$ is negligible, the probability to have a radius $R$ is simply (factors of
order unity will be ignored throughout in the following and $k_BT \equiv 1$)
\begin{equation}
 p(R) \propto \exp (-F_{e \ell} (R)) \Rightarrow \langle R^2 \rangle =
a^2 N,
\quad \langle F \rangle =1.
\end{equation}

In terms of a blob picture, a free Gaussian chain is a single blob as well. For
a free star polymer with $f$ arms linked to a center, we simply have
\begin{equation}
F_{e \ell} (R) = fR^2/a^2 N.
\end{equation}
The $f$ Gaussian arms do not disturb each other, each arm has a size $\langle
R^2 \rangle =a N$, and the total free energy then is $\langle F \rangle
=f$.

Let us now consider the effect of excluded volume, choosing $V=R^3$,
$\rho=fN/R^3$, to conclude $F_{\rm int} (R)= c \rho^2 V= c(fN)^2 /
R^3)$, here $c$ is related to the 2$^{\rm nd}$ virial coefficient,
\begin{equation}\label{eq_Flory}
F = F_{e \ell}(R) + F_{\rm int} (R) = f R^2 / a^2 N + c (fN)^2 /
R^3.
\end{equation}
Minimization gives
\begin{eqnarray}\label{eq_R}
\partial F/ \partial R &=& 0 \Rightarrow 2 f R/a^2N - 3 c (fN)^2 R^{-4} =
0, \nonumber \\
R&=&c^{1/5} a ^{2/5} f^{1/5} N^{3/5}.
\end{eqnarray}

This equation is in agreement with the Daoud-Cotton result based on the blob
picture when one takes for the exponent $\nu$ the Flory value $\nu=3/5$ (see
equation~(19) of Daoud and Cotton \cite{24}).
In these equations we have assumed for simplicity a constant monomer density
inside the volume taken by the star, ignoring the fact that the density near
the star core is enhanced. A more elaborate version of Flory theory takes a
radial density into account \cite{25'} but this does not change the qualitative
features of the results.

The free energy then becomes
\begin{equation}\label{eq_FE}
F_{\rm star} \propto c ^{2/5} f^{7/5} N^{1/5}.
\end{equation}

It is interesting to estimate the free energy from the Daoud-Cotton blob
picture which was not done in their paper but has been estimated by Witten and
Pincus \cite{25}, who obtained
\begin{equation} \label{eq_FEPincus}
F_{\rm star} \propto f^{3/2}\ln(N).
\end{equation}
As discussed by Johner \cite{25'}, the mean-field result Eq.~(\ref{eq_R})
actually holds for relatively weak excluded volume effects and not so strong
stretching of the arms in the star polymer. When the strength of the excluded
volume increases, a crossover to a stronger stretching is predicted with a
radius of the star polymer scaling as
\begin{equation}\label{eq_RR}
 R \propto f^{\frac{1-\nu}{2}} N^\nu, \quad \nu \approx 0.588
\end{equation}
This crossover from Eq.~(\ref{eq_R}) to Eq.~(\ref{eq_RR}) is missing when one
takes the Flory approximation for $\nu$, of course. A similar crossover has
also been pointed out for the height of planar brushes, which scales as $h
\propto \sigma_g^{1/3}N$ in the mean-field regime, yet as $h \propto
\sigma_g^{\frac{1-\nu}{2\nu}}N$ in the regime where excluded volume is very
strong ($\sigma_g$ being the grafting density \cite{Kreer,KB}). However, in
numerical work it is very difficult to study such small differences between
Eqs.~(\ref{eq_RR}) and (\ref{eq_R}), or between Eq.~(\ref{eq_FE}) and
(\ref{eq_FEPincus}) \cite{Hsu}. Note that Eq.~(\ref{eq_FE}) for the single chain
($f=1$) yields the wrong result $F \propto N^{1/5}$, already mentioned
in the introduction. 

\subsection{Confinement of linear chains}

When chains are confined by walls, one must take into account that also a
Gaussian chain does not penetrate a hard wall and  hence there exists a free
energy cost of confinement. If a Gaussian chain is confined in a spherical
cavity, it experiences a free energy cost of compression (see Eq.~7.2 of the
Grosberg-Khokhlov's book) \cite{CNL,GK}.
\begin{equation}
F_{\rm compression} \propto N(a/D)^2
\end{equation}

One can argue \cite{Casassa} that the same expression (apart from a prefactor)
also holds for confinement in a tube or between plates; note that the result
follows from constraining on the accessible positions of chain  monomers.
Now for Gaussian chains $x,y,z$ components of distance vectors are
uncoupled from each other. So the elastic contribution for confinement of
Gaussian chains is
\begin{equation} \label{eq_FE_Gauss}
F_{e \ell} (R) = N(a/D)^2 + R^2 /a^2 N 
\end{equation}
where the prefactor of the term $N(a/D)^2$ [this prefactor is not written!] is
$1/3$ of the above prefactor of $F_{\rm compression}$ for planar
confinement, and $2/3$ for cylindrical confinement. Here $R$ denotes the chain
linear dimension in direction(s) parallel to the confining walls, or along the
tube axis, respectively.

For confinement between parallel plates the volumes $V=R^2D$ and the density
$\rho=N/V=N/(R^2D)$, hence
\begin{equation} \label{eq_Flory_lin}
F_{\rm conf} = F_{e \ell} (R) + F _{\rm int} (R)=N(a/D)^2 +
R^2/a^2N + cN^2 / (R^2 D) \quad ({\rm plates})
\end{equation}
From $\partial F_{\rm conf} / \partial R=0$ we find (the first term
$N(a/D)^2$ does not contribute!)
\begin{equation}
 2 R/(a^2 N) - 2 c N^2/R^3 D=0,  \quad R^4=c a^2 N^3 /D, \quad
R=c^{1/4} a^{1/2} D^{-1/4} N^{3/4}
\end{equation}

Again, we conclude that the Flory argument provides the correct scaling of $R$
with $D$ and $N$, as suggested by the blob picture \cite{GK} and stressed
already in various previous papers \cite{Brochard,Sung}.

For confinement in a tube, the volume is $V=RD^2$, the density $\rho=N/V=N/(R
D^2)$ and hence
\begin{equation}
F_{\rm conf} = F_{e \ell} (R) + F_{\rm int} (R) = N(a/D)^2
+ R^2 /a^2 N + cN^2 / (R D^2) \quad , \quad ({\rm tube})
\end{equation}
and $\partial F_{\rm conf}/ \partial R=0$ then yields
\begin{equation}
 2 R/(a^2 N) - c N^2 /R^2D^2 = 0\,\, , \, \, R^3 = ca^2N^3/D^2, \,
R=c^{1/3} N(a/D)^{2/3}
\end{equation}

Again, the formula for $R$ agrees with the blob picture result \cite{GK}, but the
confinement free energy that Flory theory predicts is {\em incorrect}. The blob
picture yields (note $g=(D/a)^{1/\nu}$ where $g$ monomers are in a blob of diameter
$D$, $F_{\rm conf}=n_{\rm blob}=N/g$)
\begin{equation}
F^{(\rm blob)}_{\rm conf} =N(D/a)^{-1/\nu} = N (D/a)^{-5/3} \quad \quad
({\rm plates\; or\; tubes})
\end{equation}
in both the plates and tubes cases. This result is not reproduced by Flory
theory, which rather yields
\begin{equation}\label{eq_plates}
F^{(\rm Flory)}_{\rm conf} =N(a/D)^2 + c^{1/2} N^{1/2} (D/a)^{-1/2} \quad
\quad ({\rm plates})
\end{equation}
and
\begin{equation}\label{eq_tubes}
F^{(\rm Flory)}_{\rm conf} = N(a/D)^2 + c^{2/3} N (D/a)^{-4/3}\quad ({\rm
tubes})
\end{equation}
even though the radius $R$ in lateral direction shows the proper scaling. In
fact, the competition between the exponents $-2$ and $-1/2$ may lead to an
effective exponent. The fact that one cannot rely on Flory prediction for the
free energy of confined chains has occasionally been noted before
\cite{Brochard,Sung}. Note that some authors (e.g., \cite{Sung}) did not include
the therm $N(a/D)^2$ in the elastic part of the free energy,
Eq.~(\ref{eq_FE_Gauss}), and thus missed the corresponding term in
Eqs.~(\ref{eq_plates}),~(\ref{eq_tubes}). 

\subsection{Confined Stars}

Combining the results of the previous sections, we immediately obtain
\begin{equation}
F_{e \ell}^{\rm conf. star} (R) = fN (a/D)^2 + f R^2 / a^2N
\end{equation}
and, since for confinement in a slit we have $V=R^2D$, $\rho=f
N/V= f N/R^2 D$, we obtain $f^{conf. star}_{\rm int} (R)=c(fN)^2 /R^2 D$.

Using $F_{\rm conf. star} (R) = F^{conf. star.}_{e \ell} (R) +
F^{conf. star}_{\rm int} (R) \quad , \quad \partial F_{conf. star} (R)/\partial R=0$
we get
\begin{equation}
 2 fR/(a^2 N) - 2c (fN)^2 /R^3 D=0, \quad R^4=c fN^3/D,
\end{equation}
and hence
\begin{equation}
 R=(cf)^{1/4} (D/a)^{-1/4} N^{3/4}   \quad \quad ({\rm plates})
\end{equation}

This result is in perfect agreement with equation~(III-20) of Halperin and 
Alexander's paper\cite{halperin87} with respect to the powers of all 4
parameters $c,f, D/a$ and $N$. This result also can be interpreted as the
behavior of a two-dimensional star of blobs of diameter $D$, as pointed out by
Benhamou {\em et al.} \cite{19'}.

For confinement in a tube, we would predict ($V=RD^2$, $\rho=fN/V=fN/(RD^2)$)
\begin{equation}
F_{conf. star} (R) = fN(a/D)^2 + f R^2/ (a^2N) + c(fN)^2 /(RD^2)
\end{equation}
and hence
\begin{equation}
 3F_{conf. star} (R) / \partial R=0 \Rightarrow  \,\, 2 Rf /(a^2N) - c (fN)^2
/R^2 D^2 =0
\end{equation}
and finally $R^3 = c f N^3 (a/D)^2,\quad R=(cf)^{1/3}
N(a/D)^{2/3} \quad ({\rm tubes})$.

While the blob picture of Halperin and Alexander yields a confinement free
energy
\begin{equation}
F^{\rm blob}_{conf. star} \propto f Nc^{1/3} (a/D)^{5/3} \quad,
\end{equation}
the above treatment gives instead again a competition of terms $D^{-2}$,
$D^{-1/2}$.
\begin{equation}
F^{\rm Flory}_{conf. star} \propto fN (a/D)^2 + c^{1/2} f^{3/2} (D/a)^{-1/2}
N^{1/2}
\quad ({\rm plates})
\end{equation}

It is of interest to compare the latter expression to the numerical
self-consistent field results which is done in the next section.

\section{Self-Consistent Field Theory}
\label{sec_SCFT}

\subsection{Methodology}
\label{subsec_SCF_method}

The central quantity in the self-consistent field  (SCF) approach is the
mean-field free energy, which is expressed as a functional of the volume
fraction profiles and SCF potentials for all components in the system. As
explained in detail below, the calculation of the volume fraction profiles and
SCF potentials requires solving SCF equations  iteratively and numerically,
which necessarily involves space discretization, i.e. use of a lattice. In the
present work, we employ the method of Scheutjens and Fleer\cite{fleer93} which
uses the segment diameter $\sigma$ as the size of the cell; throughout this work
all the distances are reported in units of the cell size $\sigma$. A
three-dimensional version of the SCF theory is formulated on a cubic lattice in
Cartesian coordinates. 

The star polymer is modeled as a spherical core of radius (not diameter!)
$\sigma$ grafted with $f$ chains each of length $N$. In order to model a
spherical core on a cubic lattice, we use the previously detailed
method,\cite{wijmans94} whereby the volume fraction of the core segments,
$\phi_{c}(x,y,z)$ is set to unity for the lattice sites located completely
inside the core and is set equal to the volume fraction of the (cubic) lattice
site located within the sphere for the sites lying on the surface of the core.
The grafted polymers are modeled as chains of $N$ monomers all of which are of
the same size, with every monomer occupying one lattice site. Grafting is
achieved by pinning one end-segment of each of the chains on the spherical
core. 

In addition to the core and polymer segments (labeled $c$ and $p$,
respectively), the system under study also contains (monomeric) solvent and the
surface atoms, labeled $v$ and $s$, respectively. The surface atoms are placed
in the $x-y$ planes for $z\leq z_l$ and $z\geq z_u$; their location is
``frozen'', i.e. $\phi_{s}(x,y,z)=1$ for the $z$ ranges indicated above and is
equal to zero otherwise. The degree of confinement of the star polymer can be
varied by changing the width of the planar slit, $D=z_u-z_l$. Finally, the
solvent particles fill all the vacant lattice sites not taken by other species,
i.e. SCF calculations are carried out under the incompressibility constraint,
meaning that for each lattice site $(x,y,z)$ the sum of the volume fractions of
all the species must be equal to unity: $\sum_i\phi_{i}(x,y,z)=1$, where $i=c,
p, s,$ and $v$. 

The SCF approach operates with two key quantities -- density distributions of
all the species and the corresponding self-consistent field
potentials.\cite{fleer93} Within our three-dimensional SCF formalism, the
density profiles of all the species depend on three Cartesian coordinates, $x,
y$, and $z$. In the spirit of a mean-field approach, the potentials represent
average interactions of a test molecule with all the other molecules in the
system. As such, they depend on the density distributions of the molecules
which, in turn, are determined by the potentials. Thus, both quantities must be
computed simultaneously and self-consistently via a numerical iterative solution
of the SCF equations. In practice, starting from the initial guesses for the
volume fraction profiles of all the species, $\phi_{i}(x,y,z)$, one obtains the
potentials as follows:\cite{fleer93} 
\be
\beta u_i(x,y,z)=\beta u^{\prime}(x,y,z)+
\sum_{j}\chi_{ij}(<\phi_{j}(x,y,z)>-\phi_{j}^{b}),
\label{uxyz}
\ee
where $\phi_{j}^{b}$ is the bulk volume fraction of the component $j$,
$\beta=1/k_BT$, $k_B$ is the Boltzmann constant, and $T$ is the temperature. The
first term, which is called the excluded volume potential $u^{\prime}(x,y,z)$,
originates from the fact that the SCF equations are solved under the
incompressibility constraint specified earlier.
The Lagrange multiplier associated with the incompressibility constraint is
related to the excluded volume potential $u^{\prime}(x,y,z)$. The second term in
Eq.~(\ref{uxyz}) accounts for the energy of the contact interactions between the
segments of species $i$ and all other species present in the system, with
$\chi_{ij}$ being the corresponding Flory-Huggins interaction parameter. Lastly,
the angular brackets in Eq.~(\ref{uxyz}) denote the averaging of the volume
fraction profiles over the nearest-neighbor sites according to:
\be
<\phi_{i}(x,y,z)>=\frac{1}{6}\sum_{x^{\prime},y^{\prime},z^{\prime}}
\phi_{i}(x^{\prime},y^{\prime},z^{\prime})
\label{phiave}
\ee
where
$(x^{\prime},y^{\prime},z^{\prime})=(x+\alpha,y+\beta,z+\gamma)$, with
$\alpha,\beta,\gamma=-1,1$. We note that for a cubic lattice considered here,
the step probabilities to go from one lattice layer to the neighboring one
are all equal to 1/6 in $x, y$, and $z$ directions. 

Once the SCF potentials are obtained from Eq.~(\ref{uxyz}), one computes the
Boltzmann factors associated with these potentials, $G_i(x,y,z)=\exp(-\beta
u_i(x,y,z))$, which enter the calculation of propagators that are necessary to
obtain the volume fraction profiles of the polymeric species;  the volume
fraction profiles for the monomeric species are directly proportional to the
corresponding Boltzmann factors.\cite{fleer93} The volume fraction profiles thus
obtained are compared with the initial guesses substituted into
Eq.~(\ref{uxyz}), and the procedure is repeated until self-consistency between
input and output profiles is achieved within the desired numerical accuracy. 

\subsection{Confinement of Single chains: Comparison with Monte Carlo
Results}
\label{subsec_SCF_MC}

Since the above numerical version of SCF equations deals with polymer chains
described by walks on a cubic lattice, where a monomer occupies a lattice site,
and neighboring effective monomers along the chain are a lattice spacing apart,
the description is reminiscent of the treatment of polymers as self-avoiding
walks on the simple cubic lattice. The distinction, of course, is that in the
self-avoiding walk (SAW) model the condition that every lattice site 
can be occupied by a single
monomer only, holds strictly while the SCF equations enforce this condition as a
statistical average only. Thus, while in the standard SAW model (as studied by
extensive Monte Carlo simulation, see e.g. Hsu {\em et al.}\cite{Hsu2} 
for work on single chains confined in cylindrical tubes of diameter $D$) the
energy cost when two monomers sit on top of each other is infinite, the
potential of the SCF equations (Eq.~(\ref{uxyz})) implies a finite enthalpy cost
only. Similarly, in Flory theory the constant $c$ in
Eq.~(\ref{eq_Flory}),~(\ref{eq_Flory_lin})~-~(\ref{eq_tubes}) is also  related
to this finite excluded volume strength.

\begin{figure}[htb]
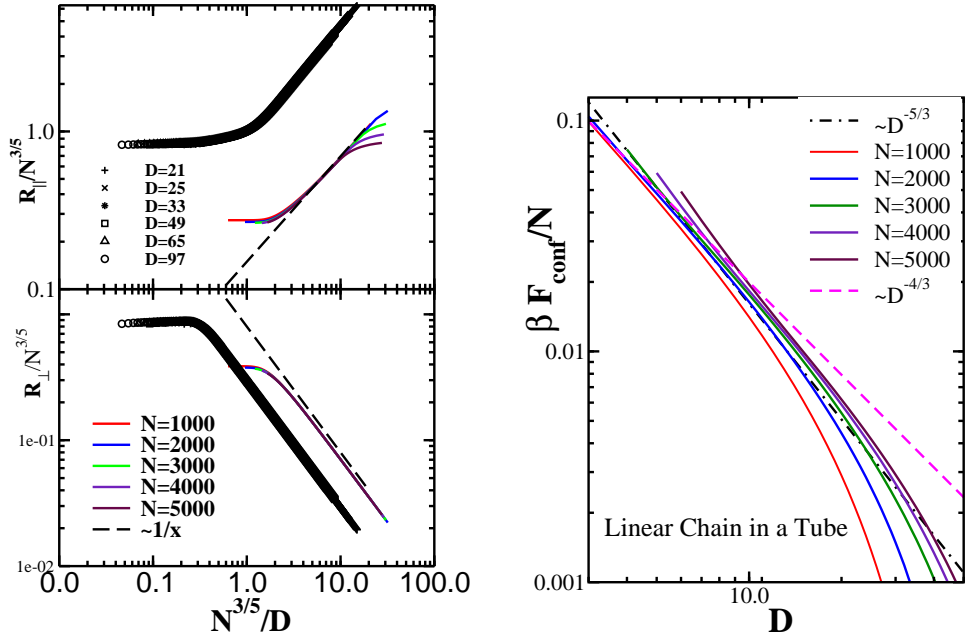

\begin{center}
\includegraphics[scale=0.33]{fig1a.eps}
\hspace{0.5cm}
\includegraphics[scale=0.33]{fig1b.eps}
\vspace{0.7cm}
\caption{(left) Comparison of MC results (symbols) for the longitudinal and
transverse components of a linear chain confined in a tube of
diameter $D$ \cite{Hsu2} with data from SCF calculations (lines). (right)
Variation of the free energy of a linear polymer with tube diameter $D$.
Theoretical power laws $~ D^{-4/3}$ and $D^{-5/3}$ (from Flory theory
and blob model, respectively) are included for comparison.
\label{fig_SCF_MC_tube}}
\end{center}
\end{figure}

Nevertheless, it is illuminating to carry out a direct comparison of Monte Carlo
data for chains confined in a cylindrical tube with corresponding SCF results
(Fig.~\ref{fig_SCF_MC_tube}a). Plotting $R_{||}/N^{3/5}$ and $R_\bot/N^{3/5}$
vs. $N^{3/5}/D$ one sees a similar quality of scaling, i.e. collapse of the data
on master curves, for both approaches. But the SCF results, are shifted to
smaller linear dimensions in comparison with the Monte Carlo data, and also the
crossover to the ``string of blob''-behavior occurs for larger values of
$N^{3/5}/D$ only. This discrepancy is easily interpreted: the mean-field
description of self-avoidance in terms of Eq.~(\ref{uxyz}) is clearly weaker
than the strict SAW condition of the Monte Carlo model. However, it is very
plausible, that the SCF equations capture the qualitative features of excluded
volume effects correctly, as far as chain linear dimensions are concerned.
However, SCF results for the free energy excess due to the confinement perform
less well (Fig.~\ref{fig_SCF_MC_tube}b). In particular, scaling predicts that $
\beta F_{conf} / N \approx const$ and simply varies with $D$ as
$D^{-1/\nu} \approx D^{-5/3}$ as long as $D$ is small in comparison to the
radius of the free chain, but this is not seen: even for $N$ as large as
$N=5000$, the data still depend on $N$ for small $D$. This fact is interpreted
by the
following consideration: Hsu et al. \cite{Hsu2} and Klushin et al.
\cite{Klushin} defined the number of blobs $n_b$ as $n_b=N(D/a)^{-1/\nu}$, they
did not allow for a prefactor (of order unity) in this relation ($a$ is the
lattice spacing of the simple cubic lattice). Then they showed that the
end-to-end distance $R_{||}$ and the free energy $F$ (in units of $k_BT)$ both scale
linearly with $n_b$ and estimated the prefactors $A$ and $B$ in the relations
\begin{equation} R_{||} = A D n_b \quad , \quad F=B n_b \quad , \quad N
\rightarrow \infty
\end{equation}
as $A \approx 0.92$, $B \approx 5.33$, for the standard SAW model on the simple
cubic lattice.

However, for comparing their results with the present calculation a different
definition for the number of blobs (which we denote as $n_{\rm blob}$ in our
paper, to distinguish it from the above convention) is more useful. We write
\begin{equation}
 n_{\rm blob} =CN(D/a)^{-1/\nu}
\end{equation}
where $C$ is a constant that is fixed by the requirement (we always deal with
the limit $N \rightarrow \infty$!)
\begin{equation}
 R_{||} = Dn_{\rm blob}= C a N (D/a)^{1-1/\nu} \approx C a N
(D/a)^{-2/3} \
\end{equation}
because the physical picture that the chain is a linear string of $n_{\rm blob}$
blobs of diameter $D$ requires that the prefactor between $R_{||}/D$ and $n_{\rm
blob}$ is precisely unity. This consideration immediately yields for the SAW
model that $n_{\rm blob}=A n_b$ and hence $F=B' n_{\rm blob}$ with $B' \approx
5.79$.

While for the SAW model the above constant $C=1/A \approx1.09$, for the present
SCF calculations it is much smaller, namely $C\approx 0.314$, reflecting the
fact that SCF represents a much weaker strength of excluded volume, as noted
above (it could be modeled by a Monte Carlo simulation where crossing of chains
at the same lattice sites is not strictly forbidden but only requires a finite
energy penalty $\varepsilon_{\rm ex}/k_BT$). For $D=10$, $N=10^3$ we hence find
that $n_{\rm blob} \approx6.8$ and hence we should have $F/N=B' n_{\rm blob}/N
\approx 0.039$ 
($B' \approx 5.79$, $n_{\rm blob} \approx6.8$, $F\approx 39$), 
which is about a factor of 3 larger than the actual
SCF result (see Fig.~1).
This consideration shows that SCF cannot be mapped onto the simulation results
by rescaling of prefactors.
 
On the basis of Eq.~(\ref{eq_tubes}), we expect that SCF theory like Flory theory does
not reproduce the blob result $F= B' CN(D/a)^{-5/3}$ but rather is described by
a competition of two terms, scaling with $D^{-2}$ and $D^{-4/3}$, respectively.
But orders of magnitude larger chains would be needed to convincingly
demonstrate that.

\subsection{SCF results for confined star polymers}
\label{subsec_SCF_res}

From the computed volume fraction profiles, one can readily obtain various
structural properties, such as the components of the radius of gyration of the
star polymer in the directions parallel ($x$, $y$) and perpendicular ($z$) to
the confining plates, e.g.
\be
R_{gx}^{2}=\frac
{\int_{-\infty}^{\infty}dx\int_{-\infty}^{\infty}dy
\int_{z_l}^{z_u}dz\phi_{p}(x,y,z)x^2}
{\int_{-\infty}^{\infty}dx\int_{-\infty}^{\infty}dy
 \int_{z_l}^{z_u}dz\phi_{p}(x,y,z)},
\label{rxsq}
\ee
where $\phi_{p}(x,y,z)$ is the volume fraction of the star polymer
segments, while $z_l$ and $z_u$ are the locations (along the $z$-axis)
of the lower and upper confining planes, respectively.

The corresponding numerical SCF results can be compared with the scaling theory.
As detailed in the previous Section, the latter predicts the following scaling
behavior of the radius of gyration component parallel to confining plates,
$R_{g||}$, (i.e. either $R_{gx}$ or $R_{gy}$ for our system):~\cite{halperin87}
\begin{equation}
R_{g||}\sim \sigma f^{1/4}N^{3/4}D^{-1/4},
\label{rpar}
\end{equation}
where good solvent conditions are assumed. Note that exactly the same 
scaling behavior of $R_{g||}$ is also predicted by the Flory theory.

In order to test the above prediction, we have performed SCF calculations for a
star polymer for several representative values of $f$ and $N$, the former
ranging from $6$ to $12$ and the latter ranging from $250$ to $750$. We
gradually varied the degree of confinement by changing the slit width $D$. All
the calculations were carried out under good solvent conditions, for
non-adsorbing confining walls, i.e. all the Flory-Huggins interaction parameters
were set equal to $0$.
 
\begin{figure}[htb]
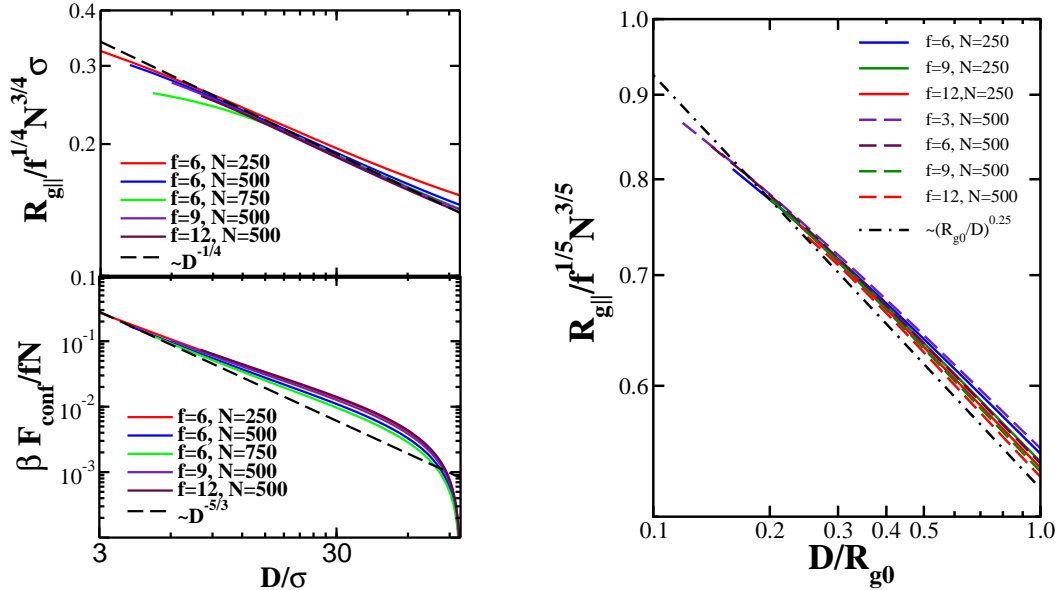

\vspace{0.7cm}
\includegraphics[scale=0.31]{fig2a.eps}
\hspace{1.0cm}
\includegraphics[scale=0.34]{fig2b.eps}
\vspace{0.5cm}
\caption{(left) Upper panel: the component of the star polymer radius of
gyration parallel to confining plates as a function of the slit width $D$ for
several values of $f$ and $N$.  Lower panel: the dimensionless confinement free
energy of a star polymer (scaled by the product $fN$) as a function of the
degree of confinement $D$ for several values of $f$ and $N$. 
SCF results are shown as solid lines, and scaling predictions are
plotted as dashed lines. 
(right) SCF results for 
$R_{g||}$ scaled by $R_{g0}$ of unconfined star (i.e., by $f^{1/5}
N^{3/5}$) plotted vs
scaled degree of confinement $x=D/R_{g0}$. According to scaling
predictions, in this form the results in the
confined regime for all $f$ and $N$ are expected to collapse onto a single curve
$~x^{-1/4}$, which is plotted as a dot-dashed line.}
\label{figscf_Egor}
\end{figure}

The summary of our results for the component of the radius of gyration parallel
to confining plates is given in the upper panel of Fig.~\ref{figscf_Egor} as a
function of the slit width $D$  for several values of $f$ and $N$. One sees that
all three scaling predictions given by Eq.~(\ref{rpar}), i.e. scaling of
$R_{g||}^{2}$ with $f^{1/2}$, $N^{3/2}$, and $D^{-1/2}$ are reasonably well
confirmed. Alternatively, in the right panel of Fig.~\ref{figscf_Egor} we
present a different plot where the ratio $R_{g||}/R_{g0}$ of $R_{g||}$ with
respect to the unperturbed gyration radius of the star-polymer, $R_{g0}$, is
scaled with the dimensionless width of the slit, $x=D/R_{g0}$, where
$R_{g0} \propto f^{1/5}  N^{3/5}$. According to Eq.~(\ref{rpar}), one should
then observe a decline by power law $\propto \left ( D/R_{g0} \right )^{-1/4}$.
As can be verified from Fig.~\ref{figscf_Egor} (right panel), the SCF results
obey the expected scaling reasonably well up to $x \approx 0.3 \div 0.4$ and
then start deviating from each other and from  the $x^{-0.25}$ line.

In addition to the radius of gyration, we have computed free energy of the star
polymer confined between two flat plates. As discussed in the previous Section,
Flory theory gives the following result for the free energy of confinement:
\begin{equation}
F_{conf. star}=fN(\sigma/D)^2+c^{1/2}f^{3/2}(\sigma/D)^{1/2}N^{1/2},
\label{fconfflory}
\end{equation}
where $c$ is related to second virial coefficient.  At the same time, scaling
theory predicts:~\cite{halperin87}
\begin{equation}
F_{conf. star}^{scaling}=fNc^{1/3}(\sigma/D)^{5/3}.
\label{fconfscaling}
\end{equation}
In other words, the blob picture predicts a scaling exponent of $-5/3$ for the
dependence of the free energy on the slit width, while Flory result is comprised
of two terms with two different scaling exponents: $-2$ and $-1/2$. 

In the lower panel of Fig.~\ref{figscf_Egor} (left), we present the SCF results
for the dimensionless confinement free energy, $\beta F_{conf}$, scaled
by the product $fN$. The free energy is plotted as a function of the plate
separation $D$ for the same pairs of values of $f$ and $N$ as in the upper
panel. One sees that while the free energy curves obtained for different values
of $f$ and $N$ collapse reasonably well onto a single master curve when scaled
by the product $fN$, this curve does not exhibit scaling behavior with $D$ that
could be convincingly described by a single scaling exponent (as predicted by
Eq.~\ref{fconfscaling}), but rather appears to follow the Flory prediction,
Eq.~(\ref{fconfflory}), which results from the competition of two terms with two
different exponents. Unfortunately, the regime where $R_{g||}$ shows the
expected power law  (for $f=9$ and $f=12,\; f=6$ shows more deviations) as 
function of $D/\sigma$, Eq.~(\ref{rpar}), is only about one decade $(6 \le
D/\sigma \le 60)$. In this regime, Eq.~(\ref{fconfscaling}) clearly is not
verified. But one decade is not enough, of course, to prove that there is a
competition of two exponents, as found in Eq.~(\ref{fconfflory}).

\section{Simulation results}
\label{sec_results}

\subsection{Molecular Dynamics details}
\label{sec_model}

We consider a three-dimensional coarse-grained model \cite{grest} of  a polymer
star which consists of $f$ linear arms with one end free and the other one
tethered to a microscopic core (seed monomer) of size $R_c$ , which is of the
order of the monomer size $\sigma$. Each arm is composed of $N$ particles of
equal size and mass, connected by bonds. Thus, the total number of monomers in
the star (excluding the core) is $fN$. The bonded interactions between
subsequent beads is described by the frequently used Kremer-Grest potential
\cite{kremergrest} $V^{\mbox{\tiny KG}}(r)=V^{\mbox{\tiny FENE}}+V^{\mbox{\tiny
WCA}}$, with the so-called 'finitely-extensible nonlinear elastic' (FENE)
potential:
\begin{equation}
 V^{\mbox{\tiny FENE}}(r)=- 0.5kr_0^2\ln{[1-(r/{r_0})^2]}
\label{fene}
\end{equation}
The non-bonded interactions between monomers are taken into account by means of
Weeks-Chandler-Andersen (WCA) interaction, i.e., the shifted and truncated
repulsive branch of the Lennard-Jones potential given by:
\begin{equation}
 V^{\mbox{\tiny WCA}}(r) = 4\epsilon\left[
(\sigma/ r)^{12} - (\sigma /r)^6 + 1/4
\right]\theta(2^{1/6}\sigma-r)
\label{wca}
\end{equation}
In Eqs.~(\ref{fene}) and (\ref{wca}), $r$ denotes the distance between the
center of two monomers (beads), while the energy scale $\epsilon$ and the length
scale $\sigma$ are chosen as the units of energy and length, respectively.
Accordingly, the remaining parameters are fixed at the values
$k=30\epsilon/\sigma^2$, $r_0=1.5\sigma$.  In Eq.~(\ref{wca}) we have introduced
the Heaviside step function $\theta(x)=0$ or 1 for $x<0$ or $x\geq 0$.
In consequence, the steric interactions in our model
correspond to good solvent conditions.

The star-polymer is placed in a slit, i.e., two parallel,  repulsive and
infinite  walls modeled by  WCA-potential of Eq.~(\ref{wca}) acting only in
$z$--direction (perpendicular to the wall). We consider slits widths from
$D=3.5\sigma$ up to $D=100\sigma$.

The equilibrium dynamics of the chain is obtained by solving the Langevin
equation of motion for the position $\mathbf r_n=[x_n,y_n,z_n]$ of each bead in
the star,
\begin{equation}
m\ddot{\mathbf r}_n = \mathbf F_n^{\mbox{\tiny FENE}} + \mathbf F_n^{\mbox{\tiny
WCA}} -\gamma\dot{\mathbf r}_n + \mathbf R_n(t), \qquad
(n=1,\ldots,fN)
\label{langevin}
\end{equation}
which describes the Brownian motion of a set of interacting monomers. In
Eq.~(\ref{langevin}) $\mathbf F_n^{\mbox{\tiny FENE}}$ and $\mathbf
F_n^{\mbox{\tiny WCA}}$ are deterministic forces exerted on monomer $n$ by the
remaining bonded and non-bonded monomers, respectively. The influence of solvent
is split into slowly evolving viscous force $-\gamma\dot{\mathbf r}_n$ and
rapidly fluctuating stochastic force $\mathbf R_n$. The random, Gaussian force
$\mathbf R_n$ is related to the friction coefficient $\gamma$ by the
fluctuation-dissipation theorem. The integration step was $0.002$ time units
(t.u.) and time is measured in units of $\sqrt{m\sigma^2/\epsilon}$, where $m$
denotes the mass of the beads, $m=1$. The ratio of the inertial forces over the
friction forces in Eq. (\ref{langevin}) is characterized  by the Reynolds number
${\rm Re} = \sqrt{m \epsilon}/\gamma \sigma$ which in our setup is ${\rm Re}=4$.
In  the course of our simulation the velocity-Verlet algorithm was employed to
integrate the equations of motion (\ref{langevin}).

Starting configurations were generated as radially straight chains fixed to the
immobile seeded (core) monomer, prior to being placed into a slit with given
width $D$. As a next step, star-polymers were equilibrated until they adopted
their equilibrium configurations in the confinement geometry. In this step sizes
and monomer density profiles have been determined. The required equilibration
time depends on $f$ and $N$, ranging from $5\times 10^5$ time steps for the smallest stars up
to $10^7$ for the largest. For star polymers with higher functionality, $f >
12$, a larger core bead diameter $\sigma_{core} > \sigma$ has been taken so as
to accommodate the $f$-grafting beads of the arms on the surface of the core
bead. An initially random placement of these beads on the surface of the core
bead has been subsequently rendered to produce a uniform separation between
the grafting points (as if being subjected to repulsive Coulomb potential), using
 the 'diffuse.cpp' code, developed by J. D. Lettvin \cite{Lettvin}.

In order to sample the interaction force of a trapped star, for each
equilibrated configuration (at a given width $D$), a production run was
performed during which time averaged force 
$\langle \mathcal{F}(t,D) \rangle\equiv \mathcal{F}(D)$
exerted by monomers on both walls as well as monomer density profiles were
calculated. The run was continued until uncertainty of the average forces  were
less than $0.01\sigma$.

\subsection{Numerical results for confined stars}
\label{subsec_MD_res}

\begin{figure}[ht]
\vspace{0.50cm}
\includegraphics[scale=0.415]{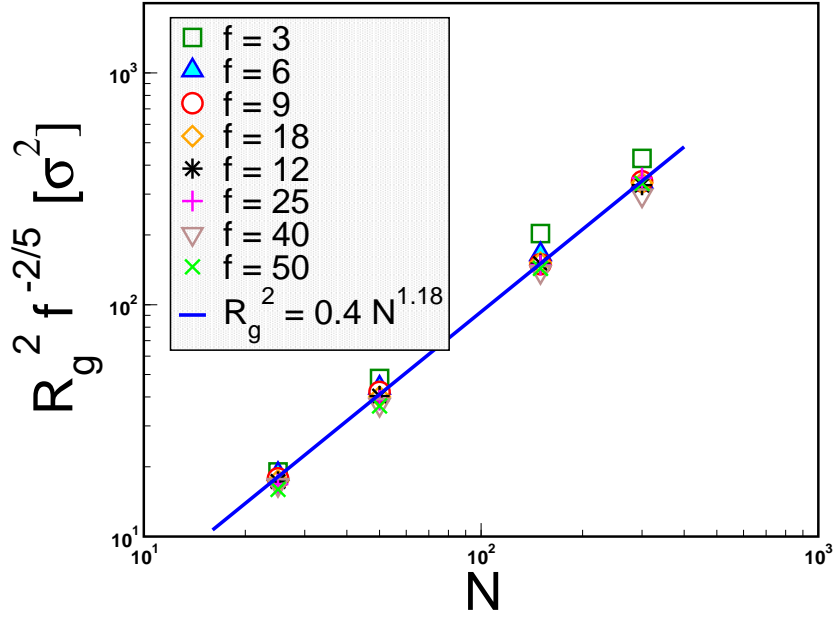}
\vspace{0.5cm}
\caption{Radius of gyration $R_g$ as a function of arm length $N$ and the
number of arms $f$ for free star-polymers (no confining walls). 
MD results are given by symbols, and scaling prediction is indicated by
the solid line.
\label{fig_Rg_free}}
\end{figure}

We begin by considering the scaling properties of free unconfined stars,
examining the scaling of the mean squared radius of gyration of the star
polymers with the number of arms $f$ and their length $N$, the
corresponding results are shown in 
Fig.~\ref{fig_Rg_free}. Since the predicted behavior is $R_{g0}^2 \propto
f^{2/5} N^{2\nu}$, we show a log - log plot of  $R_{g0}^2 f^{-2/5}$ versus $N$,
indicating a wide range of $f$-values from $f=3$ to $f=50$. It is clear that
this scaling with $f$ cannot be really expected to hold yet for the
smallest number of arms, $f=3$, and
indeed the data for $f=3$ are slightly off the master curve (which on the
log-log plot is the simple straight line corresponding to $R_{g0}^2 f^{-2/5} =
0.4 N^{1.18}$).Thus, chain lengths $25 \le N \le 300$ and number of arms $f \ge
6$ suffice to reach the scaling regime for our simple bead-spring model. 
We also note here that our scaling results for unconfined stars are in
agreement with earlier simulation work of Grest.~\cite{28}

\begin{figure}[htb]
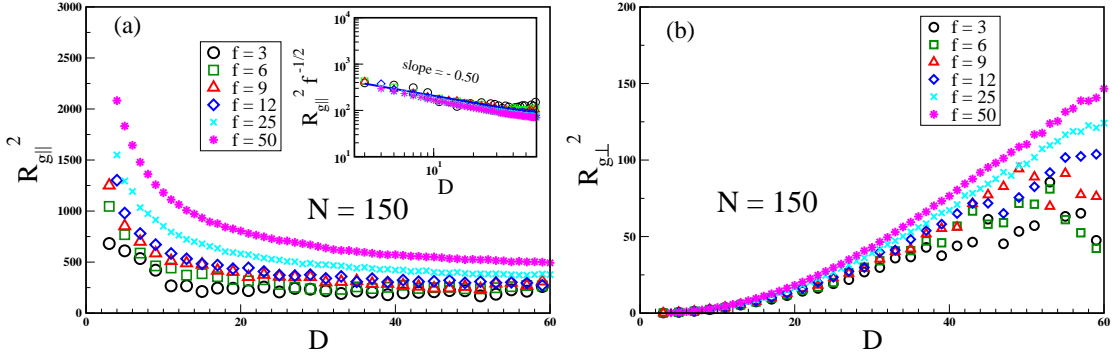

\vspace{0.7cm}
\begin{center}
\includegraphics[scale=0.27]{fig4a.eps}
\vspace{0.5cm}
\includegraphics[scale=0.27]{fig4b.eps}
\caption{Parallel (a), and perpendicular (b) components of the radius of
gyration, $R_{g||}^2, R_{g\bot}^2$, for confined star polymers with different
$f$ and $N=150$, plotted as a function of slit width $D$; MD results
are given by symbols. In the
inset a scaling plot (collapse) of data is displayed in $\log -
\log$-coordinates, indicating the predicted relationship $R_{g||}^2 \propto
f^{1/2} \left( \frac{\sigma}{D} \right)^{1/2} N^{3/2}$ (shown by a
solid line)~\cite{halperin87}.\label{fig_Rg_D}}
\end{center}
\end{figure}

Next, we present the data for the star linear dimensions confined between
walls, varying $D$ over a wide range, $3 \le D \le 60$. Now there is a need to
distinguish components $R_{g||}^2$ and $R_{g\bot}^2$ of the gyration radius
parallel and perpendicular  to the confining walls.  For the free unconfined
star (basically reached at $D=60$) we must recover $R_{g\bot}^2 = R_{g||}^2/2$
since then all three Cartesian components $R_{gz}^2 = R_{g\bot}^2$, $R_{gx}^2 = R_{gy}^2 =
R_{g||}^2/2$ are equivalent.  Fig.~\ref{fig_Rg_D} shows that with decreasing
$D$ there is a progressing decrease of $R_{g\bot}^2$ while $R_{g||}^2$ is
enhanced. Note also that for a small number of arms ($f=3 , 6$) there are still
rather pronounced statistical inaccuracies which indicate that for confined
star polymers  with few arms equilibration suffers from unexpectedly large
relaxation times. But replotting the data as $R_{g||}^2f^{-1/2}$ versus $D$ on
a log-log plot,  Fig.~\ref{fig_Rg_D} (inset), shows that we reproduce the
predicted behavior $R_{g||}^2 \propto f^{1/2}D^{-1/2}$.

\begin{figure}[ht]
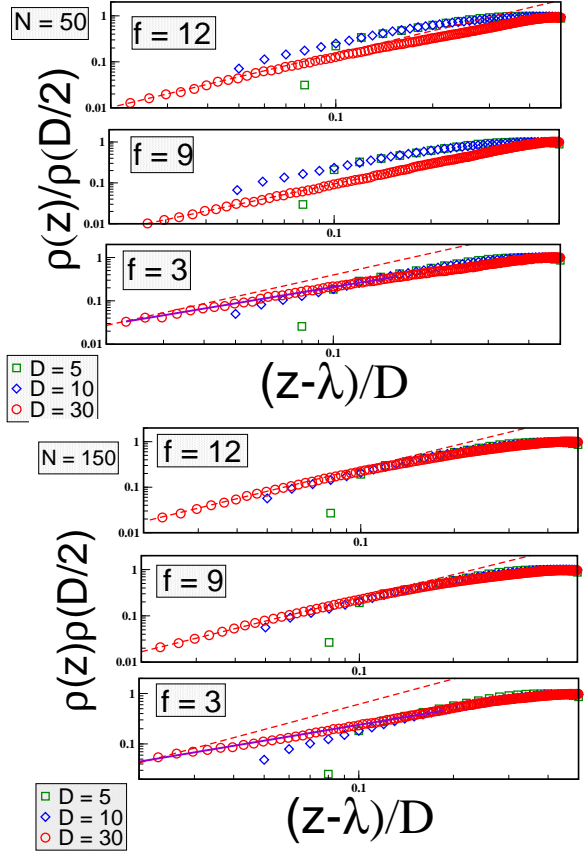

\begin{center}
\includegraphics[scale=0.27]{fig5a.eps}
 
\hspace{0.5cm}
\includegraphics[scale=0.27]{fig5b.eps}
\caption{Monomer density profiles for trapped star-polymers presented for $N=50$
[left panel] and $N=150$ [right panel] for various numbers of arms $f$ and slit
widths as indicated; MD results
are given by symbols. Dashed line (scaling prediction) corresponds to
$z^{1/\nu}$, 
full lines for $f=3$ only: $z^{1.25}$ for $N=50$ and $z \propto  z^{1.05}$ for $N=150$.
\label{fig_dens_prof}}
\end{center} 
\end{figure}

In the Flory theory it was explicitly assumed that inside the disk-like volume
taken by the confined star the density of monomers is essentially constant.
Fig.~\ref{fig_dens_prof} indicates that this assumption is a rather poor
approximation. Rather the density profile $\rho(z)/\rho(D/2) \propto
[(z-\lambda) / D]^{1/\nu}$, if $D$ is large and $z \ll D$. Here an
extrapolation length $\lambda = 0.51$ is used.

\begin{figure}[ht]
\vspace{0.9cm}
\begin{center}
\includegraphics[scale=0.43]{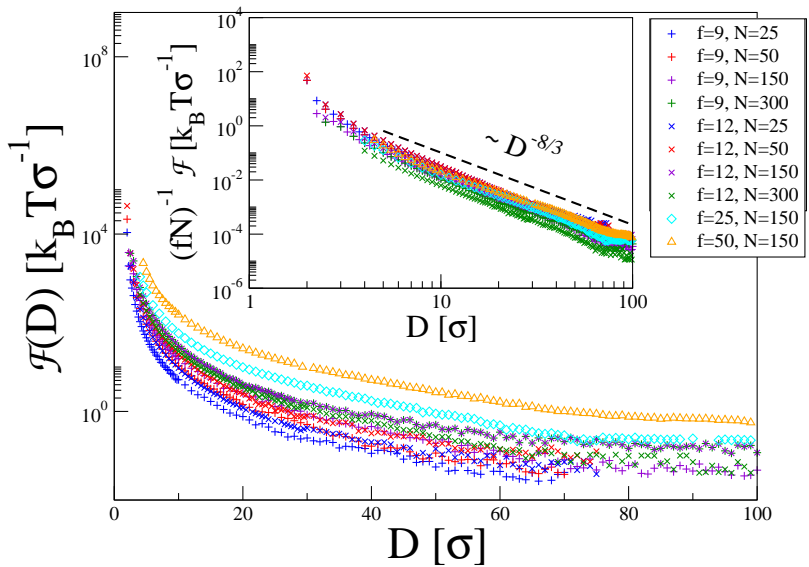}
\vspace{0.50cm}

\caption{Mean Potential Force $\mathcal{F}(D)$ felt by confined star-polymers vs
separation distance $D$ of a slit plotted  for different values of $f$ and $N$
as indicated; MD results
are given by symbols. 
The inset displays the same relationship in log-log coordinates;
dashed line gives the scaling prediction. 
\label{fig_F_D}}
\end{center}
\end{figure}

\begin{figure}[ht]
\begin{center}
\includegraphics[scale=0.43]{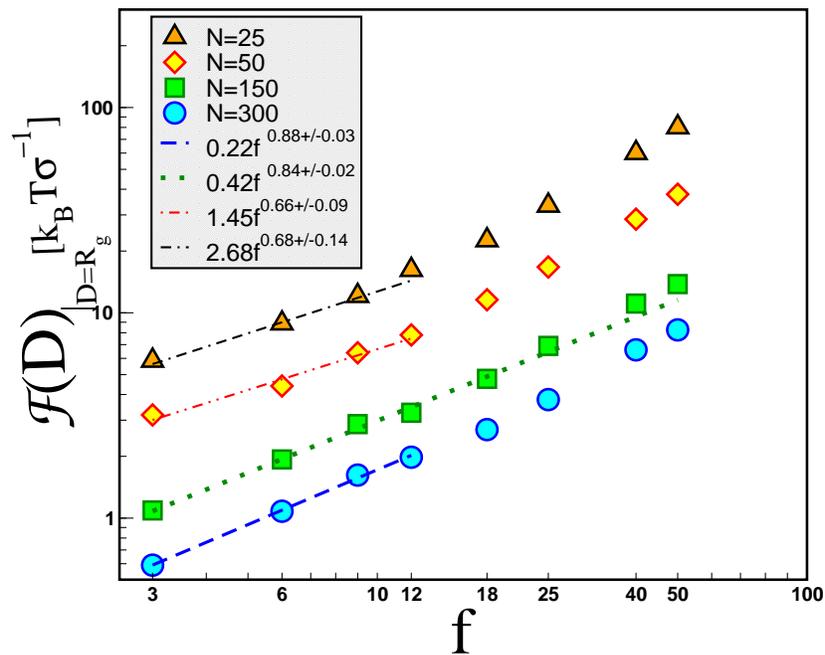}
\vspace{0.50cm}

\caption{Mean Potential Force $\mathcal{F}(D)$, felt by confined star-polymers vs
functionality $f$ in a slit of width $D = R_{g0}$, plotted  for different
values of $N$ as indicated; MD results
are given by symbols, lines are the best fits to MD data.  
\label{fig_F_f}}
\end{center}
\end{figure}

Figure~\ref{fig_F_D} shows the force $\mathcal{F}(D)$ that the monomers of the confined
star exert on the walls, choosing slits of widths up to $D = 100$. The log-log
plot in the inset shows that the data are roughly compatible with the behavior
predicted by scaling theory, $\mathcal{F}(D) \propto NfD^{-8/3}$. 
Since the scaling is not perfect, we have tested for the possible presence of correction terms
proportional to $f^{3/2}$ that one might expect when $D$ is of the order of the
linear dimension of an unperturbed star, $R_{g0}$. We note that the force of a
star interacting with a single wall is predicted to be \cite{30}
\begin{equation} \label{eq_F_1wall}
 \mathcal{F}(z) \propto f^{3/2}/z,\quad z < R_{g0},
\end{equation}
if $z$ is the distance between the star center and the constraining wall; when
$D$ is of the order of $R_{g0}$, one might expect that $\mathcal{F}(D)$ is of the same
order, i.e. 
$\mathcal{F}(D) \propto f^{3/2}/D$. To test for this hypothesis, $\mathcal{F}(D)$ for
$D = R_{g0}$ is plotted vs $f$ in Fig.~\ref{fig_F_f}. One sees, however, that
this hypothesis is not fulfilled, the $f$-dependence is much weaker.  The
reason for this problem is not clear to us.

\begin{figure}[ht]
\begin{center}
\includegraphics[scale=0.61]{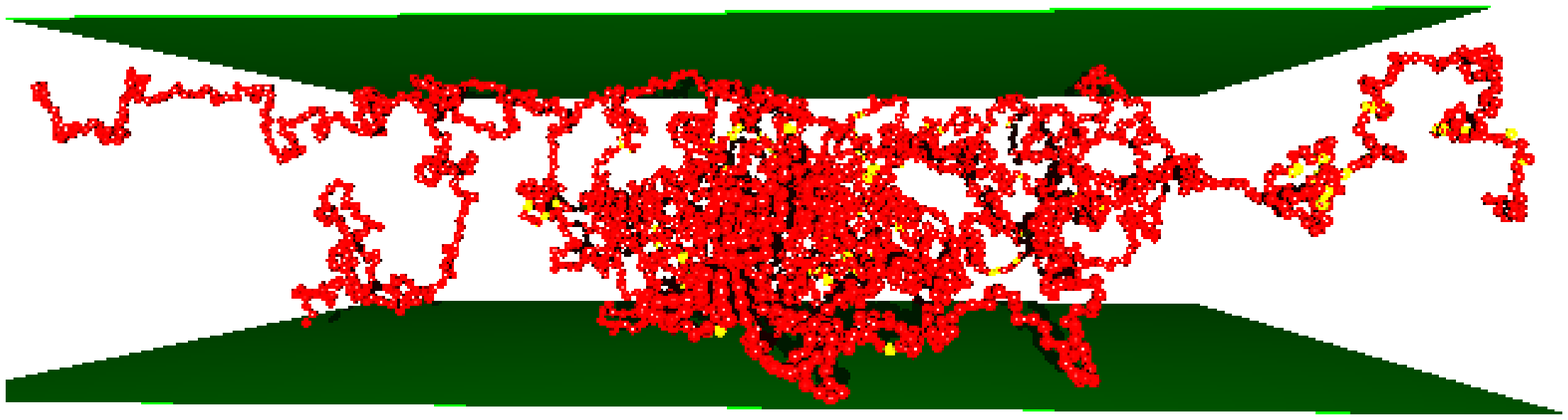}
\vspace{1.0cm}

\includegraphics[scale=0.61]{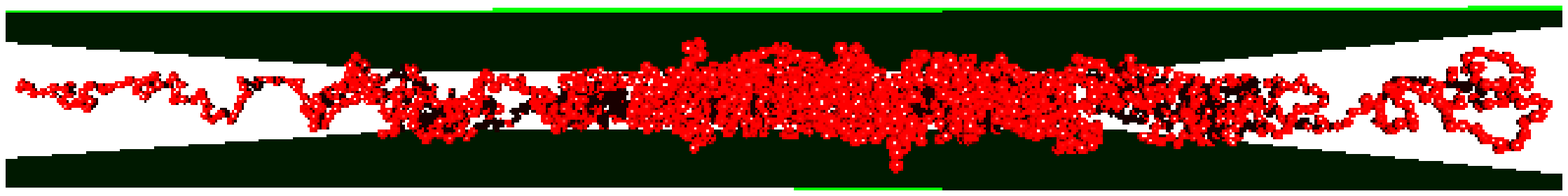}
\vspace{1.0cm}

\includegraphics[scale=0.61]{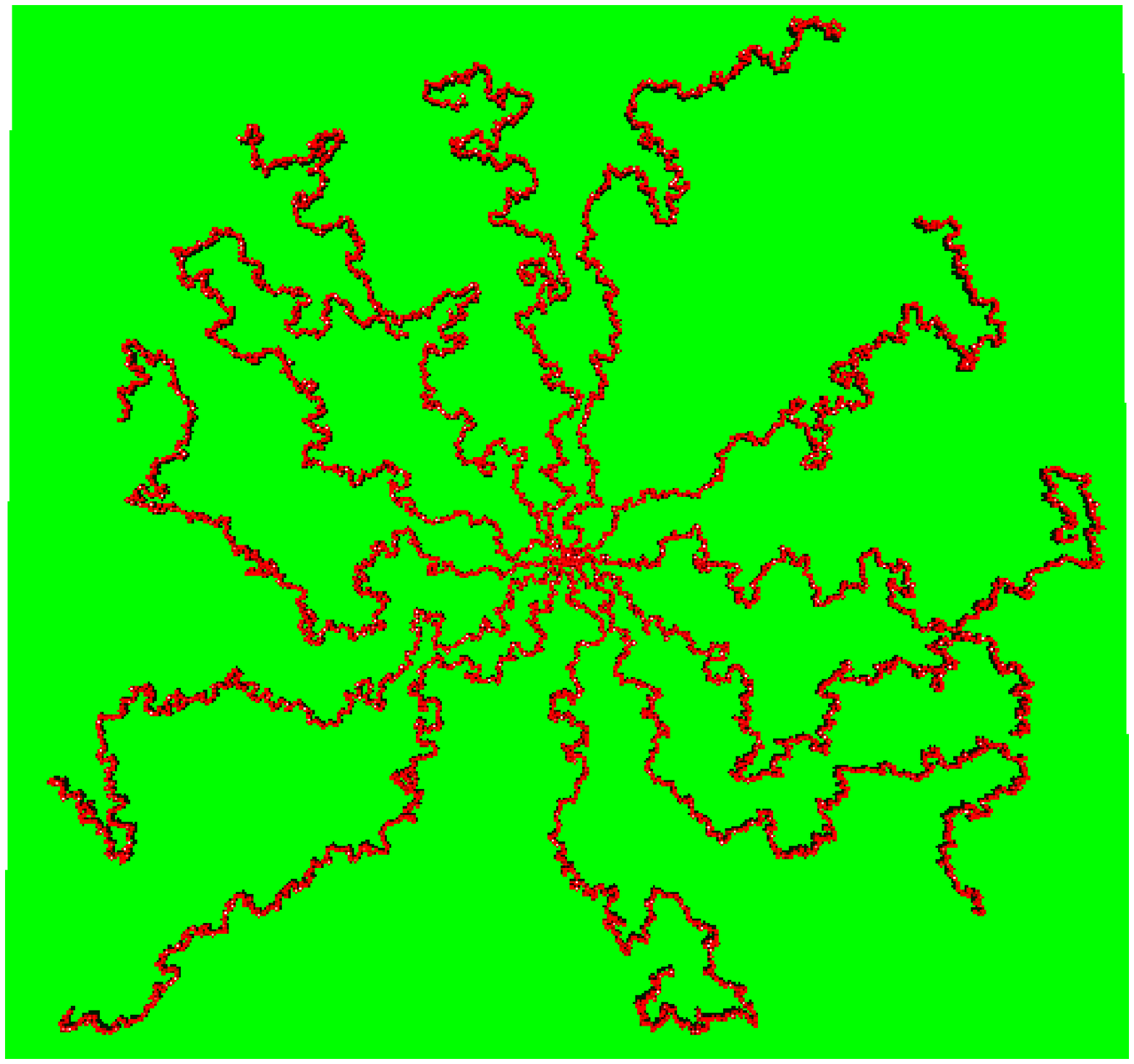}
\caption{Snapshots of a star polymer with $N = 300$ and $f = 12$ in a slit with
separation between the walls $D = 30$ (top), $D = 10$ (middle) (side views),
and $D = 5$ (bottom) (top view). \label{fig_snapshots}}
\end{center}
\end{figure}

For the strongly confined stars, on the other hand, the result 
$\mathcal{F}(D) \propto NfD^{-8/3}$ appears plausible from the snapshot picture,
Fig.~\ref{fig_snapshots}: apart from the region in the immediate neighborhood
of the center, the confined star is equivalent to an assembly of $f$
confined linear chains of length $N$, so the system contains $NfD^{-5/3}$
blobs, and this number of blobs is then just  the free energy of the confined
star.

\begin{figure}[htb]
\begin{center}
\includegraphics[scale=0.33]{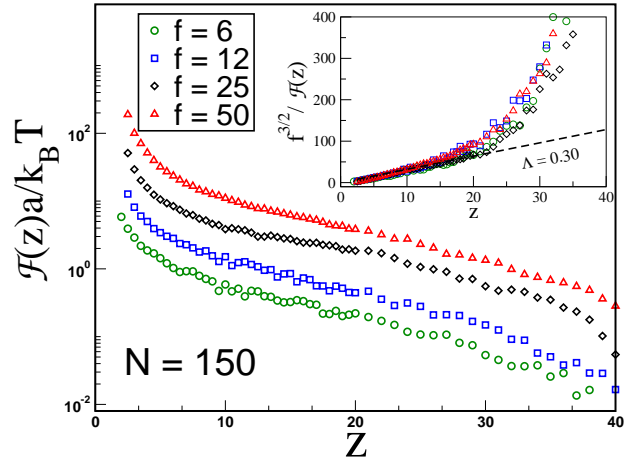}
\caption{Force exerted on a repulsive wall by a single star polymer with the
core monomer fixed at distance $z$ from the wall. Here $N=150$ and the number
of arms $f$ is given as parameter; MD results
are given by symbols. In the inset one observes a master plot
collapse of $\mathcal{F}(z)^{-1}$ for $z \le 15$, 
indicating that the force $\mathcal{F}(z)$ and
the  free energy, $F(z)$, scale $\propto f^{3/2}$ with
functionality $f$; dashed line is the scaling prediction.
\label{fig_1wall}}
\end{center}
\end{figure}

\subsection{Force between a star polymer and a repulsive wall}

As noted above, the force acting from a star polymer interacting with a wall a
distance $z$ from its center is predicted to be $\mathcal{F}(z) \propto f^{3/2}/z$, for $z
< R_{go}$ (Eq.~(\ref{eq_F_1wall})). However, our results for the force between
a star polymer interacting with two walls a distance $D$ apart never gave any
evidence for the scaling with $f^{3/2}$. Rather, we found a scaling linear in
$f$ for strong confinement (Fig.~\ref{fig_F_D}) or even weaker than linear in
$f$ for moderate confinement (Fig.~\ref{fig_F_f}).

One might suspect that our failure to see a scaling proportional to $f^{3/2}$
could be due to the problem that we are not yet in the regime where a $f^{3/2}$
scaling holds (too short or too few arms, etc.). In order to show that this is
not the case, we have studied stars in semi-infinite space, interacting with a
wall at distance $z$, for precisely the same choices of $f$ and $N$ as used in
the previous section. In practice, in our simulations a ``semi-infinite space"
also is realized by a film geometry, where one wall is at distance $z$ and the
other wall is a distance much larger than $R_{go}$, so that the configuration of
the star is not at all affected by this second wall. Fig.~\ref{fig_1wall} gives
a plot of our results for $N=150$ and $f$ in the range from $f=6$ to $f=50$. The
insert replots these data in the form of $f^{3/2}/f(z)$ versus $z$: one sees
that for $3 \leq z \le 50$ the data nicely superimpose on a straight line, in
agreement with Eq.~(\ref{eq_F_1wall}). Thus, the idea that one can easily relate
the problem confining a star with walls at distances $z=D/2$ from both sides,
simply has to be abandoned. In the latter case, there are on average always
$f/2$ arms in the region above the star center $(D/2 < z < D)$ and below it ($0
< z < D/2)$. In the case of a single wall, more and more arms move above the
star center, the closer the center approaches the wall, and hence the symmetry
of the star   with respect to the center is completely destroyed. This effect
must be responsible for the different scaling behavior with $f$.

\section{Conclusions}
\label{sec_concl}

In the paper we have presented a discussion of how confinement of star polymers
under good solvent conditions affects their linear dimensions parallel and
perpendicular to the confining repulsive parallel walls, and we considered also
the free energy cost of confinement (or the related issue of the repulsive force
exerted by the macromolecule on these walls). We have restricted attention to
the case of long arms of the star polymers (typically we choose $N=150$ in our
Molecular Dynamics simulations, but for some purposes both shorter and longer
arms, up to $N=300$, have been considered, while for the self-consistent field
work somewhat longer chains, from $N=250$ to $N=750$, could be used). The number
of arms $f$ was varied from $f=6$ to $f=50$, avoiding the regime where
properties are dominated by monomeric crowding near the star center. Although
for testing asymptotic scaling relations it would be desirable to be able to
study even significantly longer chains, we stress the fact that our work
compares well to the regime that would be experimentally accessible (recall that
in our coarse-grained model every effective monomer corresponds to a group of
about 3 to 5 chemical monomers along the backbone of a real polymer chain
molecule.)

We find that both self-consistent field calculations and Molecular Dynamics
results confirm a result originally proposed by Halperin and
Alexander
 
\cite{halperin87} that radius of the confined star scales with $f, N$ and the
distance $D$ between the walls as $R \propto f^{1/4} D^{-1/4} N^{3/4}$, which can
be interpreted in terms of the behavior of a two-dimensional star of blobs of
diameter $D$. The corresponding confinement free energy $F_{\rm conf}
\propto f N (a/D)^{5/3}$, or force on the walls, $\mathcal{F} \propto f N(a/D)^{-8/3}$ is
compatible with the MD results, while there seems to be some systematic
disagreement with both self-consistent field calculations and the even simpler
Flory theory approach. While the latter predicts correctly the scaling of the
linear dimensions, this success to some extent is fortuitous, and the free
energy is incorrectly predicted, as in the case of the single unconfined chain.
We argue that self-consistent field calculations for such confinement problems
share some of these problems of Flory theory, despite their much more elaborate
character. To elucidate this problem further, we have also performed
calculations of linear chains (with $N$ up to $N=5000)$ confined in cylindrical
tubes, and compared them to corresponding Monte Carlo calculations. 
In this context we draw attention to a comparison of self-consistent field calculations 
and Flory theory with simulations of spherical brushes confined in spherical 
cavities \cite{cerda}. They also find 
that these theories in general fail to reproduce the 
proper scaling behavior.  

While in the general case the theory \cite{halperin87} predicts for the
confinement free energy a competition of several terms, some of them scaling
with $f$ like $f^{3/2}$, and a similar scaling also applies for the force $\mathcal{F}(z)$
between a star polymers at distance $z$ from a wall \cite{30}, $\mathcal{F}(z)
\propto f^{3/2}/z$, a relation which we also nicely confirm
(Fig.~\ref{fig_1wall}), we do not find any evidence for a regime with a
$f^{3/2}$ scaling in the confinement free energy (or force $\mathcal{F}(D)$, respectively,
see Fig.~\ref{fig_F_f}). Presumably, regimes of much larger $N$ and larger $f$
are necessary to elucidate all the different regimes proposed by theory
\cite{halperin87}.

\underline{Acknowledgement}: One of us (A.M). thanks the Deutsche
Forschungsgemeinschaft (DFG) for support under grant No Bi 314/23, another
(S.A.E.) thanks both the DFG (grant No SFB 625/A3) and the Alexander von
Humboldt foundation for partial support. We are grateful to H.-P.Hsu for
providing us with her data (from Ref. \cite{Hsu2}) for comparison in
Fig.~\ref{fig_SCF_MC_tube}.
Computational time on the PL-Grid Infrastructure is gratefully acknowledged.

\end{document}